\documentclass{iaus}
\usepackage{graphicx}

\title[VVDS: NIR Follow-Up] 
{NIR Follow-Up of the VVDS 02hr Field
}

\author[Temporin et al.]   
{S. Temporin$^1$, A. Iovino$^1$, H. J. McCracken$^{2,5}$, M. Bolzonella$^3$,\\ M. Scodeggio$^4$, 
and the VVDS Team\thanks{The complete author list, as well as bibliographic references
can be found in the Appendix.}}

\affiliation{$^1$INAF-Osservatorio Astronomico di Brera, Via Brera 28, 
20121 Milano, Italy 
\\[\affilskip]
$^2$ Institut d'Astrophysique de Paris, UMR 7095, 98 bis Bvd Arago, 75014 Paris, France\\[\affilskip]
$^3$INAF-Osservatorio Astronomico di Bologna, Via Ranzani 1, Bologna, Italy 
\\[\affilskip]
$^4$INAF-IASF, Via Bassini 11, Milano, Italy \\[\affilskip]
$^5$Observatoire de Paris, LERMA, 61 Avenue de l'Observatoire, 75014 Paris, 
France}

\pubyear{2006}
\volume{235}  
\pagerange{?? -- ??}
\date{?? and in revised form ??}
\jname{Galaxy Evolution across the Hubble Time}
\editors{F.Combes \& J. Palous, eds.}
\begin{document}

\maketitle

\begin{abstract}
We present a new K-band survey covering 623 arcmin$^2$ in the VVDS 0226-0430 
deep field down to a limiting magnitude K$_{\rm{Vega}}$ $\leq$ 20.5. 
We use the spectroscopic sample extracted from this new K-band
catalogue to assess the effectiveness of optical-near infrared color selections in 
identifying extreme classes of objects at high redshift. 
\end{abstract}

\firstsection
\section{The VVDS F02 K-selected sample}
Near-infrared (NIR) selected samples are ideal tools for studying the process of
mass assembly at intermediate/high redshift, thanks to the advantages of NIR with 
respect to optical selection. Within the context of the VIMOS-VLT Deep Survey
(see Garilli \etal, Lamareille \etal, and Vergani \etal, this volume, 
for additional information on the VVDS), we present a new K-band survey in the
VVDS 0226-0430 deep field (F02), already covered by the purely 
flux-limited VVDS spectroscopic survey (17.5 $\leq$ I$_{\rm{AB}}$ $\leq$ 24; 
\cite[Le F\'evre \etal\ 2005]{olf05}). Deep ancillary photometric data in this 
field are available through the VVDS (BVRI; \cite[McCracken \etal\ 2003]{hjmcc03})
and the CFHT Legacy Survey ($u^{\ast}g^{\prime}r^{\prime}i^{\prime}z^{\prime}$).
The new K-band data, obtained with SOFI at ESO-NTT, extend those described in
\cite{ai05}, thus covering a total contiguous area of 623 arcmin$^2$ to a
limiting magnitude K$_{\rm{Vega}}$ $\leq$ 20.5 (90\% completeness). 
Our photometric catalogue includes 8857 objects down to K$_{\rm{Vega}}$ 
$\leq$ 20.25. Galaxy counts are in good agreement with those from the literature.
The angular correlation function does not show any peculiarity as a function 
of magnitude and angular scale and is broadly in agreement with results from 
the literature.
The K-selected spectroscopic sample contains 1792 galaxies 
with good quality redshifts. A minimal incompleteness in color 
arises at K$_{Vega}$ $>$ 19.8, as the reddest objects are disfavoured 
by the I$_{\rm{AB}}\,\leq$ 24 limit.  For this red tail of the galaxy 
color distribution we rely on the wide multiwavelength coverage to obtain 
good quality photometric redshifts following the method adopted in \cite{oi06}. 
For this purpose, we note that the use of K-band data reduces 
considerably the percentage of catastrophic errors (see Appendix).
A detailed description of this K-band sample is given in \cite{st07a}.
Since the VVDS survey is purely flux-limited, this sample is ideal for assessing the 
effectiveness of different optical-NIR color selections to identify 
extreme classes of objects like extremely red objects (EROs) and high-redshift 
galaxies (BzK; \cite[Daddi \etal\ 2004]{daddi04}). Some of our first results 
are summarized in the Appendix, while a 
thorough discussion of the various color selections is presented in 
\cite{st07b}.

\appendix
\section{On-line material}

\subsection{Photometric redshifts}

\begin{figure}
\centering
\includegraphics[height=10cm]{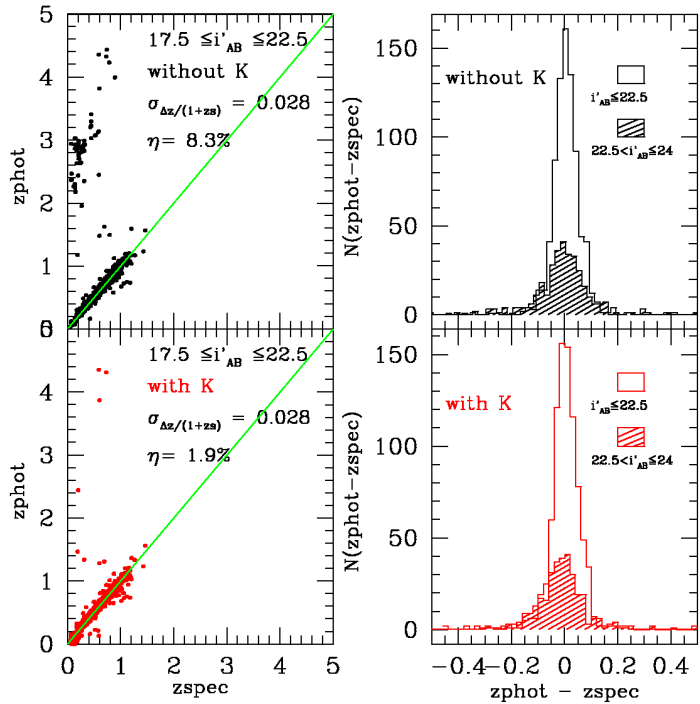}
\caption{Comparison between spectroscopic and photometric redshifts obtained
with and without the use of the K-band. Distributions of the difference
between photometric and spectroscopic redshifts are shown for two magnitude
ranges.}
\label{fig:photoz}
\end{figure}

Photometric redshifts have been derived for the whole K-band catalogue
by using the code Le Phare (developed by S. Arnouts and O. Ilbert; 
http://www.lam.oamp.fr/arnouts/LE\_ PHARE.html), following the method
described in \cite{oi06}. The use of photometric redshifts in our case
is important especially for the red tail of the galaxy color distribution,
which is missed by the spectroscopic survey because of the imposed magnitude
limit (I$_{\rm{AB}}\,\leq$ 24). For these objects the redshift distribution
of the spectroscopic sample cannot be introduced as an a priori information
in the probability distribution function. In this case, the quality of
the photometric redshifts is significantly improved by the additional 
information coming from the K-band data. In Fig.~\ref{fig:photoz} 
we show a comparison between spectroscopic redshifts and photometric redshifts 
obtained with and without the use of the K-band data. 
It appears clearly that the use of the K-band reduces considerably the 
fraction of catastrophic errors $\eta$.
Further details are given in \cite{st07a}.

\subsection{EROs and high redshift objects in the K-selected spectroscopic sample:\\ 
Color selections}

\begin{figure}
\centering
\hbox{
\includegraphics[width=0.52\textwidth]{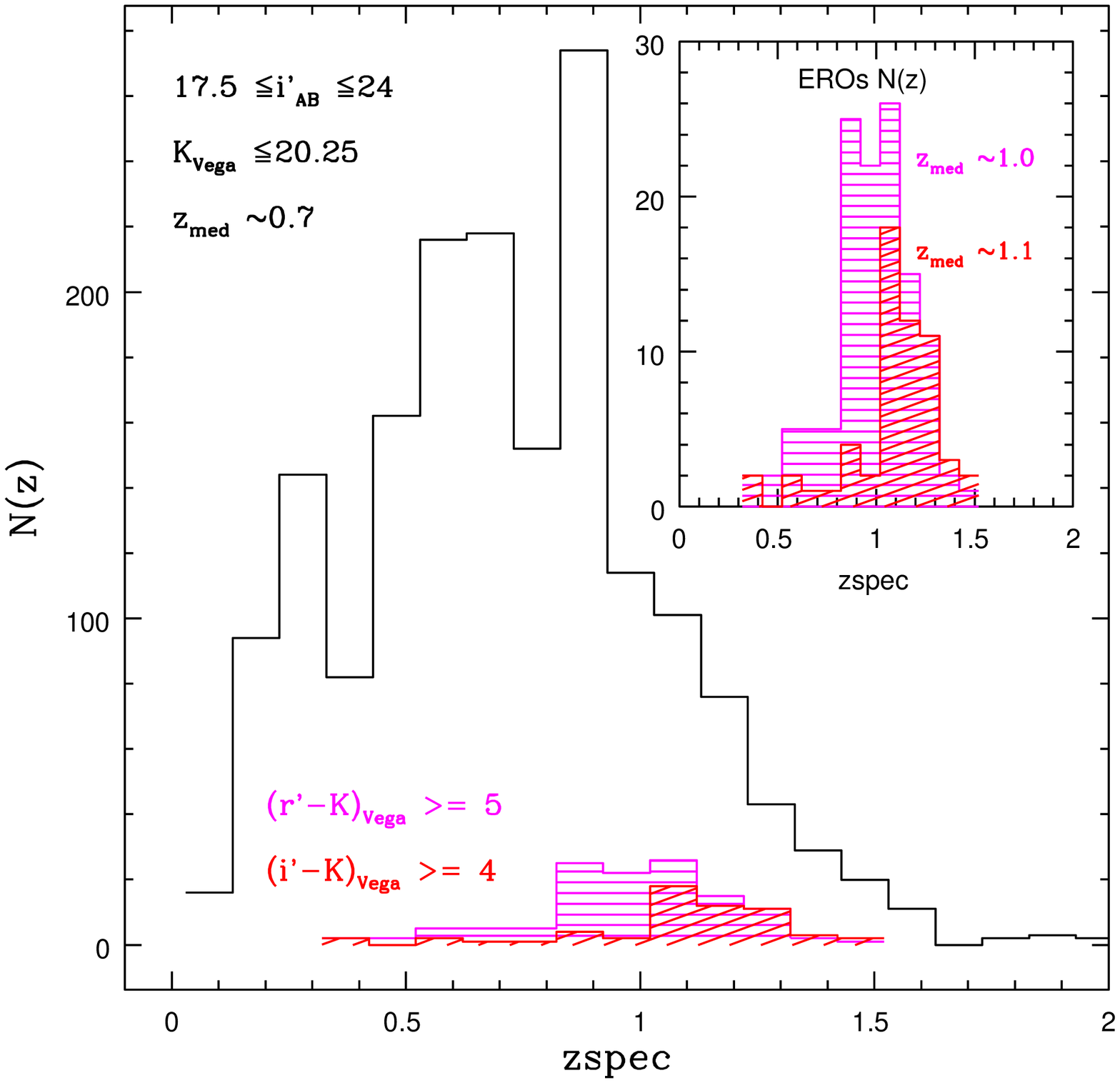}
\hskip -1.5in
\vbox{
\includegraphics[width=0.5\textwidth]{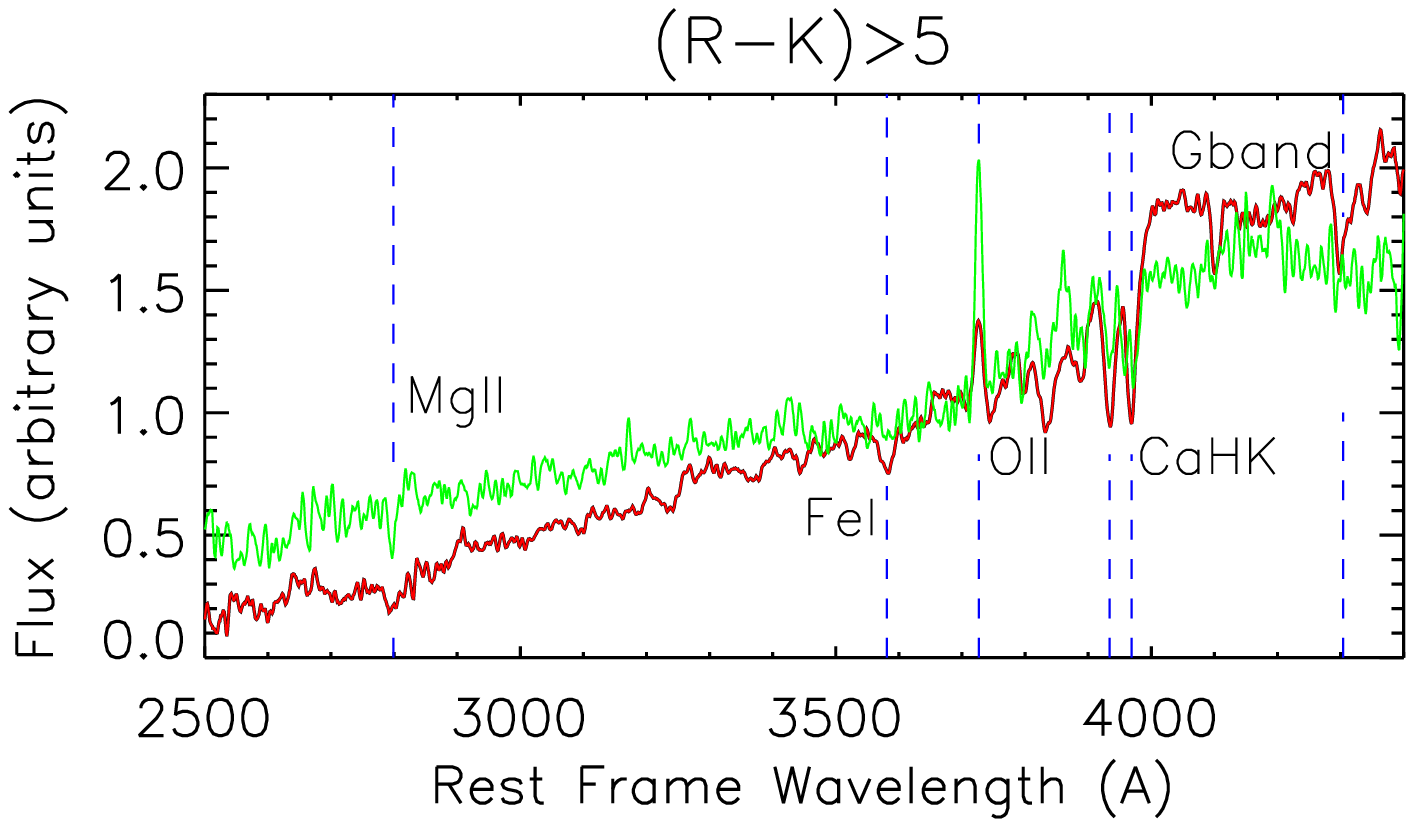}
\includegraphics[width=0.5\textwidth]{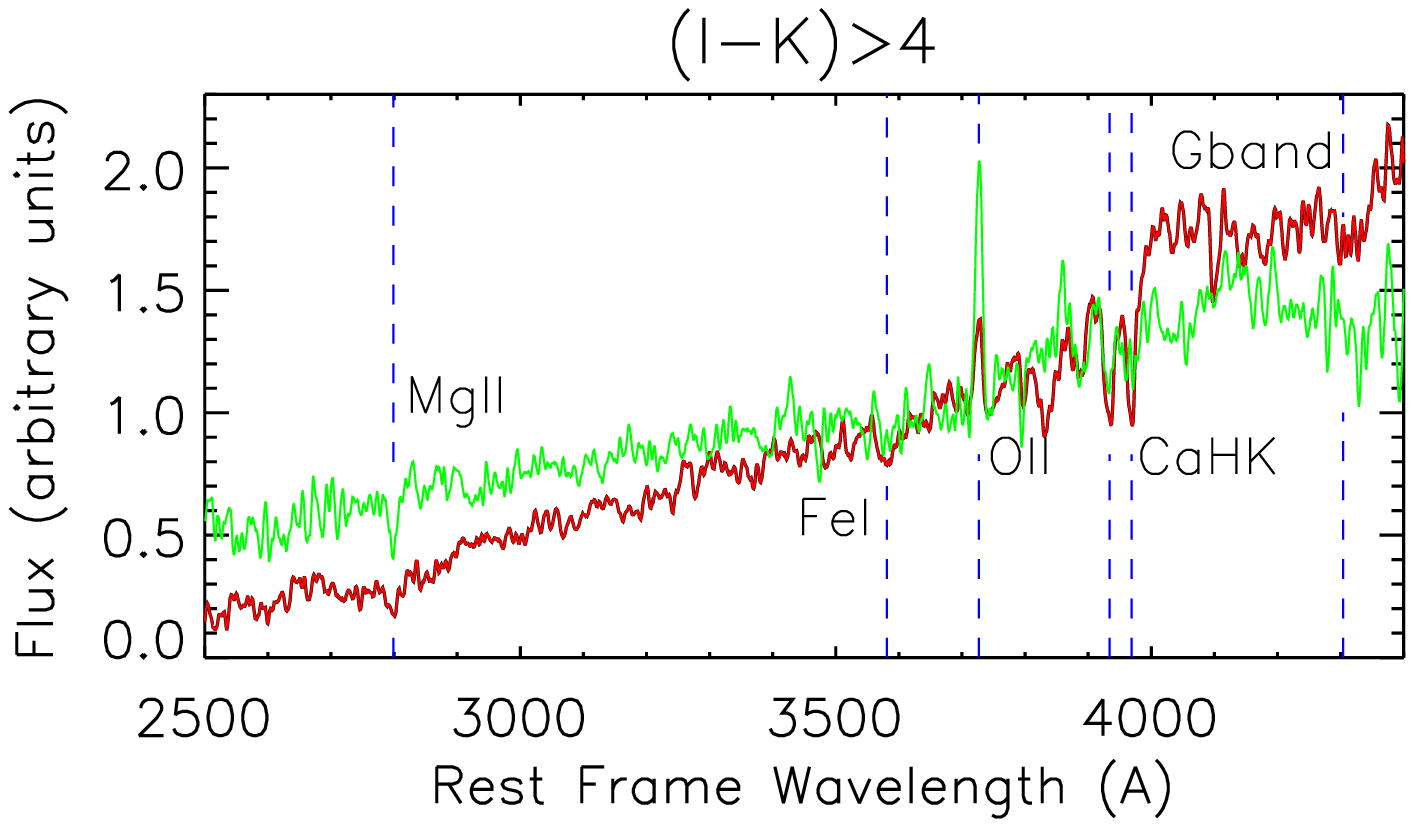}}}
\caption{\emph{Left.} Redshift distributions of the whole spectroscopic 
sample (black), the (r$^{\prime}$-K)-EROs sample (magenta), and the (i$^{\prime}$-K)-EROs sample
(red). \emph{Right.} Composite spectra of the ``early''-type (red) and
``late''-type (green) galaxies extracted from the two EROs samples.}
\label{fig:eros}
\end{figure}

Within our K-selected spectroscopic sample we have identified EROs 
according to two color selections, 
(r$^{\prime}$-K)$_{\rm{Vega}}$ $>$ 5 and (i$^{\prime}$-K)$_{\rm{Vega}}$ $>$ 4. Our 
spectroscopic sample includes 148 and 70 EROs with secure redshifts that satisfy 
the 1st and 2nd criterion, respectively, down to K$_{\rm{Vega}}$ $\leq$ 20.25. 
The whole spectroscopic sample 
has a median redshift z$_{\rm{med}}$ $\sim$ 0.7, while 
EROs have z$_{\rm{med}}$ $\sim$ 1.0 - 1.1 (Fig.~\ref{fig:eros}), in 
agreement with other samples from the literature. 
The color (i$^{\prime}$-K) appears more effective than (r$^{\prime}$-K) in selecting galaxies 
within a narrow redshift range, up to z$\sim$1.4.

A first comparison of the rest-frame observed spectral energy distributions 
(SEDs) with \cite{cww80} 
templates suggests that both samples of EROs are 
dominated by early-type galaxies. 
This is confirmed by the best-fit models resulting from the SED fitting with 
the code Le Phare. Spectra of both samples, accordingly divided into two 
broad classes (early and late), were used to build the composite spectra. 
The ``early''-type composite spectra built from 96 (r$^{\prime}$-K)-EROs and 
54 (i$^{\prime}$-K)-EROs, show spectral features typical of early-type galaxies. 
The ``late''-type composite spectra built from 17 (r$^{\prime}$-K)-EROs and 15
(i$^{\prime}$-K)-EROs, show signs of active star formation (Fig.~\ref{fig:eros}). 
An extended discussion is contained in \cite{st07b}.

\subsection{High redshift objects in the K-selected spectroscopic sample:\\ 
Effectiveness of the BzK Color selection}

\begin{figure}
\centering
\hbox{
\includegraphics[width=0.55\textwidth]{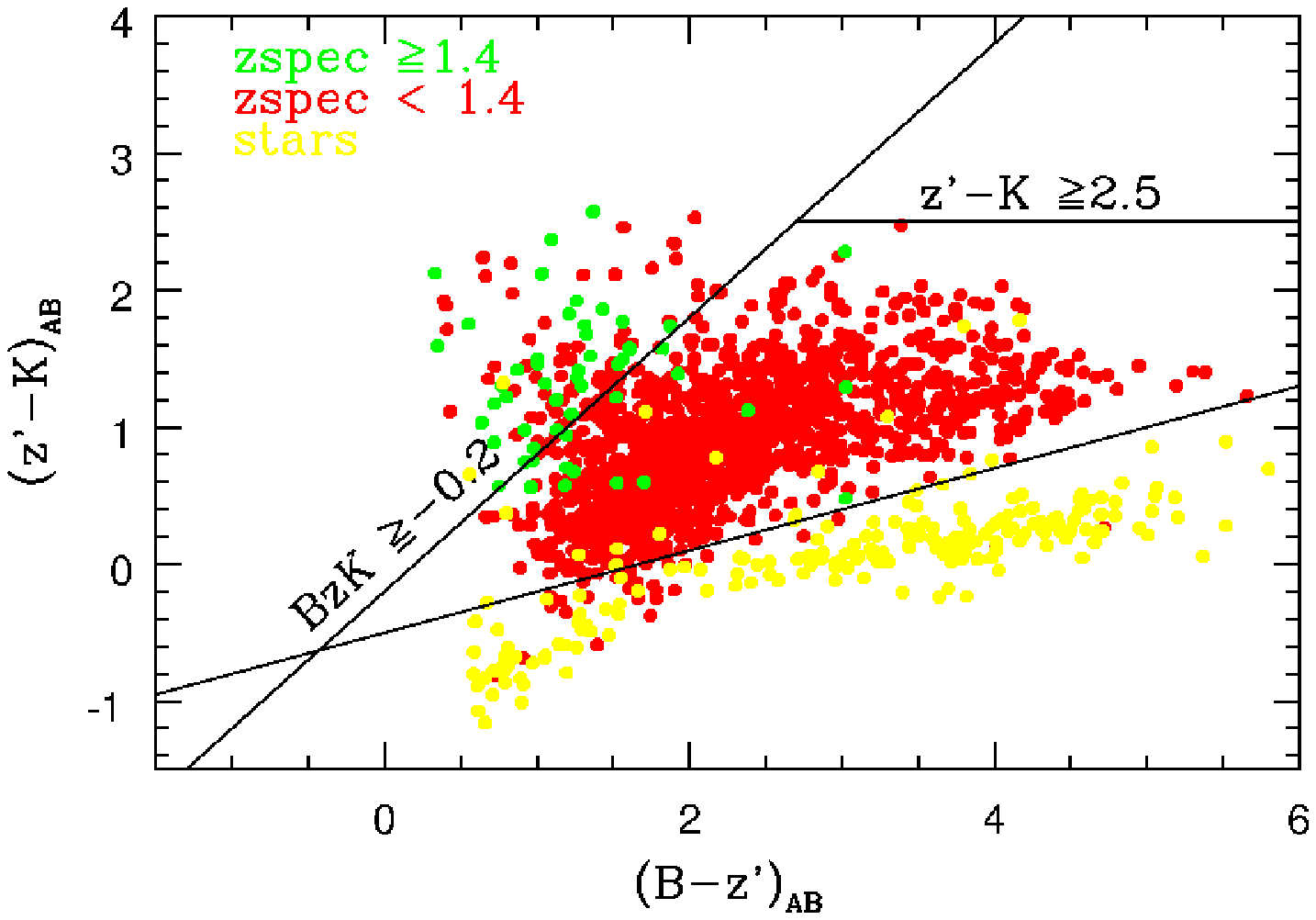}
\includegraphics[width=0.38\textwidth]{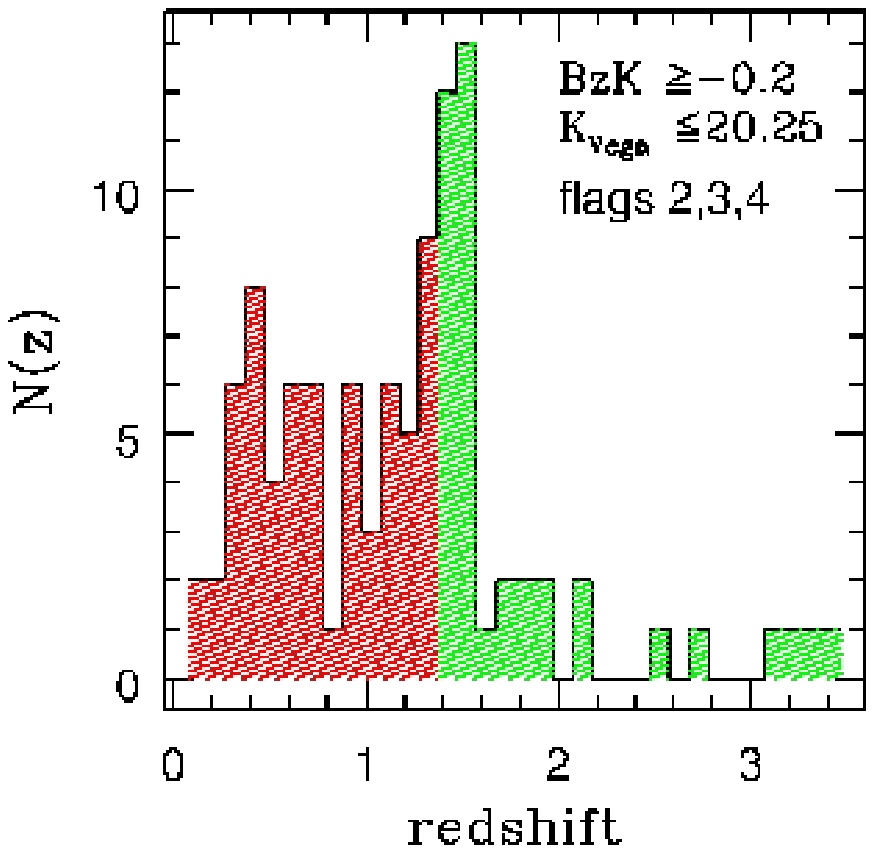}}
\hbox{
\includegraphics[width=0.35\textwidth]{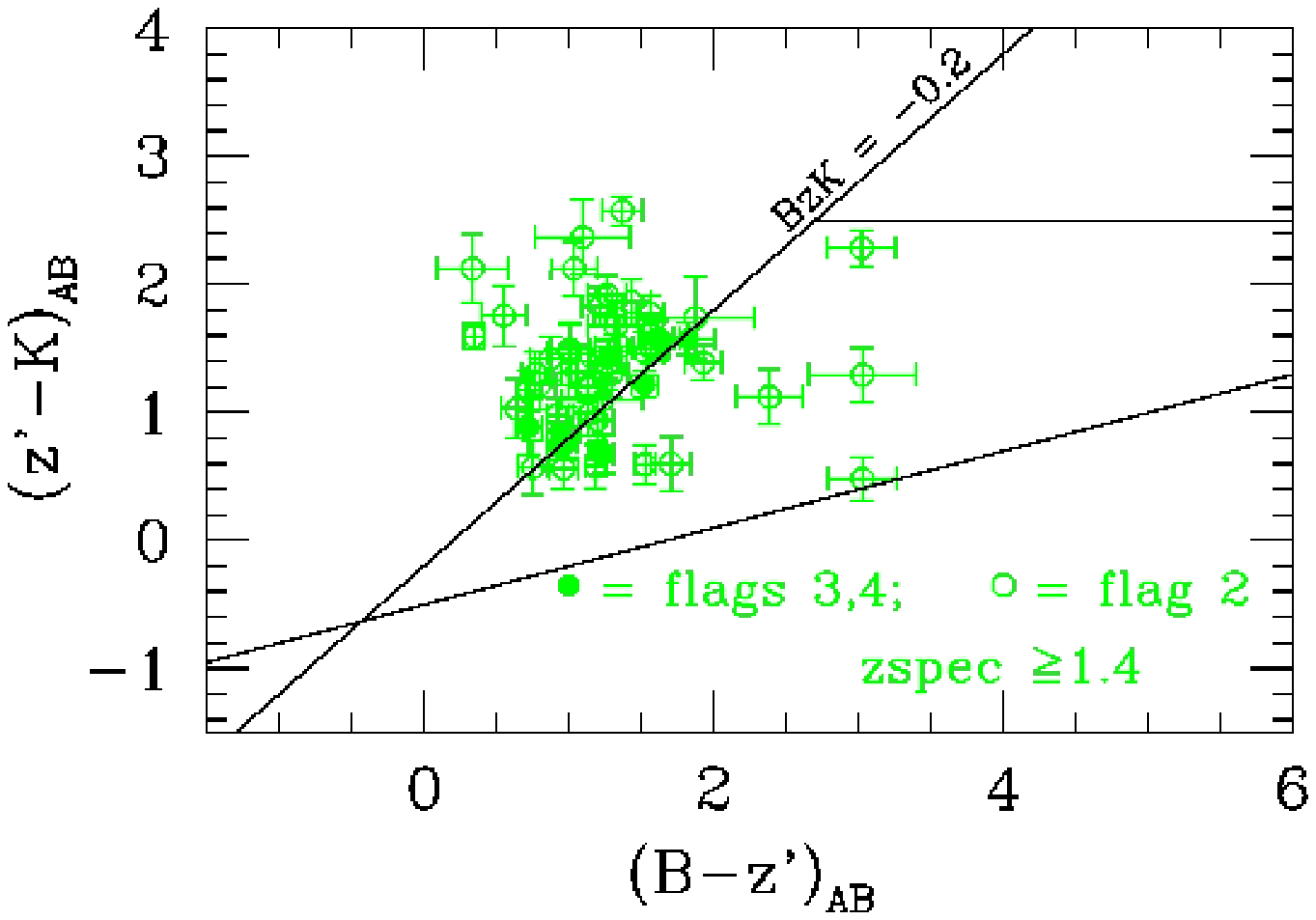}
\includegraphics[width=0.35\textwidth]{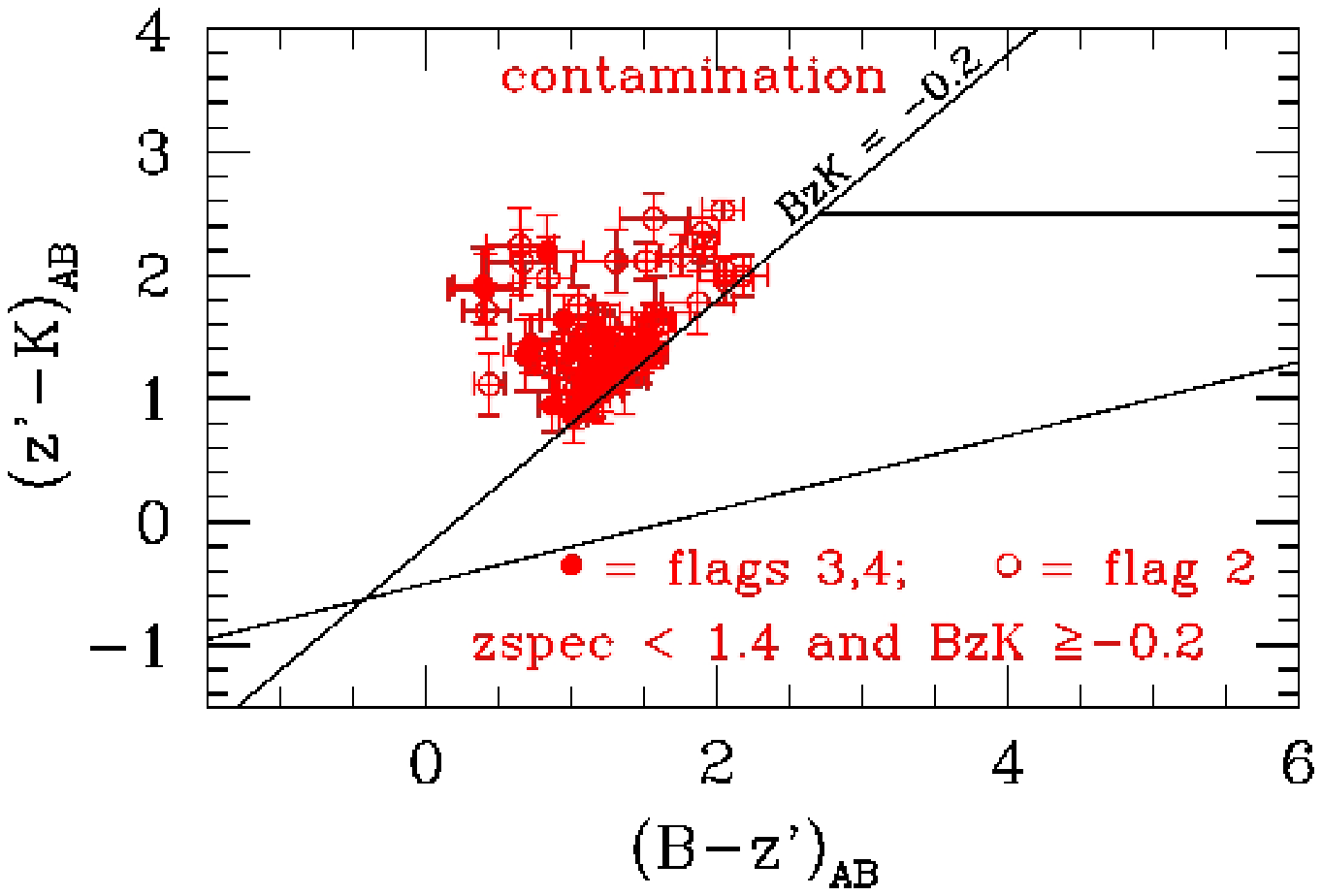}
\includegraphics[width=0.25\textwidth]{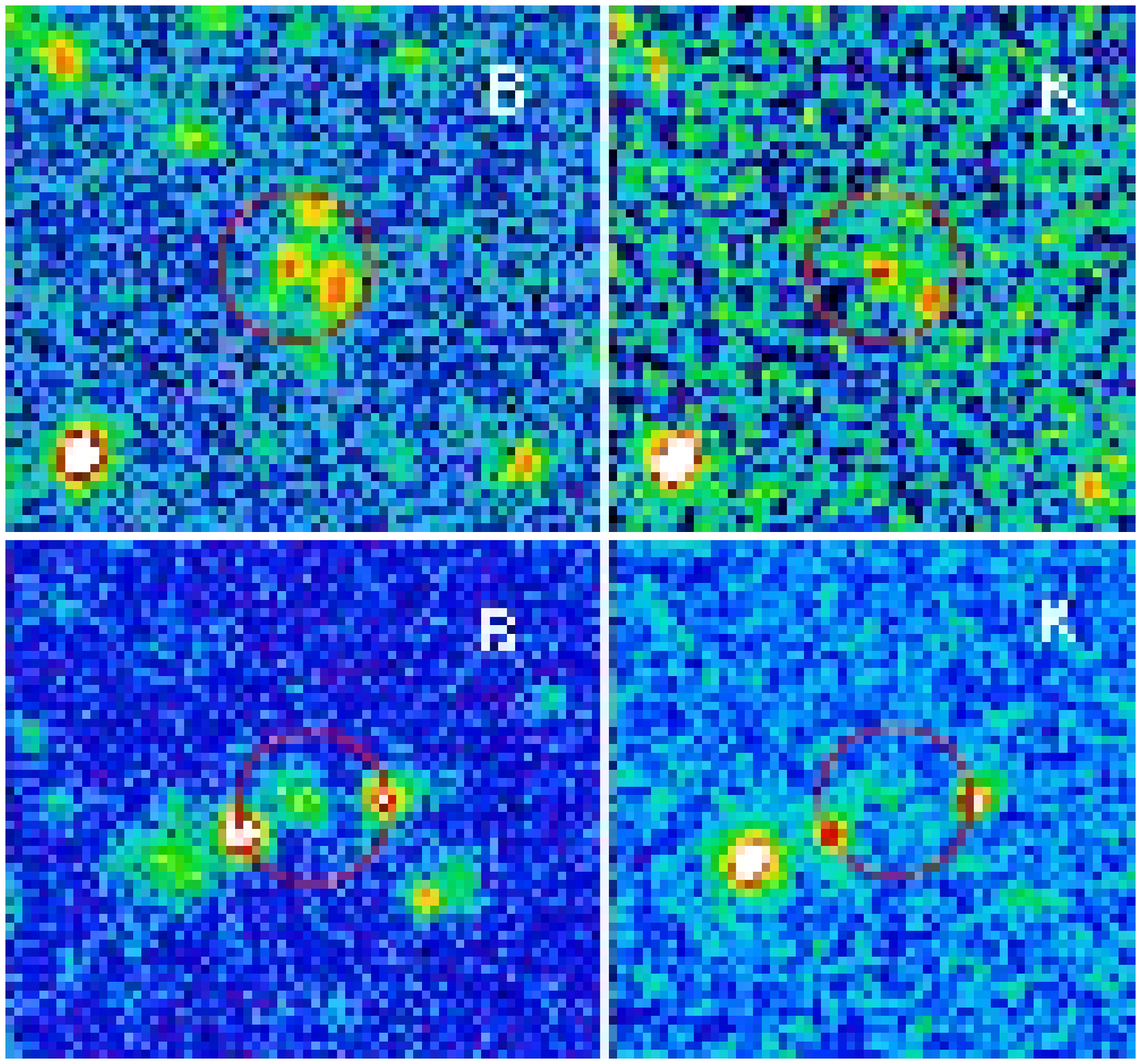}}
\caption{\emph{Upper left.} BzK diagnostic diagram for the K-selected
spectroscopic sample. \emph{Upper right.} Redshift distribution of
galaxies with BzK $\geq \, -$0.2. \emph{Lower left} BzK diagrams
for galaxies with z $>$ 1.4, irrespective of their colors (green), 
and for galaxies with z $<$ 1.4 and BzK $\geq \, -$0.2 (red).
\emph{Lower right.} Example of B and K-band images of low-redshift
galaxies in multiple systems, which contaminate the high-redshift
locus of the BzK diagram. The red circle has a radius of 3 arcsec.}
\label{fig:bzk}
\end{figure}

We explored the effectiveness of the BzK diagnostic diagram 
(\cite[Daddi \etal\ 2004]{Daddi04})
in selecting high redshift galaxies from our
K-selected spectroscopic sample.
Our filter set was verified
to be similar to the one used by \cite{Daddi04}, hence no color
term was applied.
Although the BzK method succeeds in identifying the 
high-redshift galaxies in our spectroscopic sample, we 
found a 64\% contamination by low-redshift galaxies in 
the relevant area of the diagram, BzK = (z$^{\prime}$$-$K) $-$ (B$-$z$^{\prime}$) $>\,-$0.2,
 which is expected to be populated by star-forming 
galaxies at redshift z $>$ 1.4 (Fig:~\ref{fig:bzk}). 
True high-redshift galaxies and low-redshift contaminants are 
shown separately in Fig.~\ref{fig:bzk}. 
Error bars are from SExtractor (\cite[Bertin and Arnouts 1996]{ba96})
measurements.
Apparently, photometric errors do not justify the observed contamination. Also, 
in all three bands involved, the magnitude distribution of the contaminant galaxies 
and that of high redshift galaxies are similar. 
 
Interestingly, an inspection of the images of these objects revealed that 
$\sim$ 55\% of the contaminants are members of tight (projected) pairs or 
multiple systems. Two such examples are shown in Fig.~\ref{fig:bzk}.
The contamination rate lowers to 44\% when tight pairs or 
multiple systems like those in Fig.\ref{fig:bzk}, for which photometry is 
necessarily less reliable, even in our sub-arcsec 
seeing images, are discarded. 
We note that the I$_{\rm{AB}}$ = 24 magnitude limit of the spectroscopic
survey, which disfavours high-redshift and particularly red objects,
might play a role in producing a higher contamination than usually 
observed (\cite[e.g. $\sim$ 20\% in the spectroscopic sample of Daddi \etal\ 2004]{Daddi04}).

\begin{figure}
\centering
\hbox{
\includegraphics[width=0.55\textwidth]{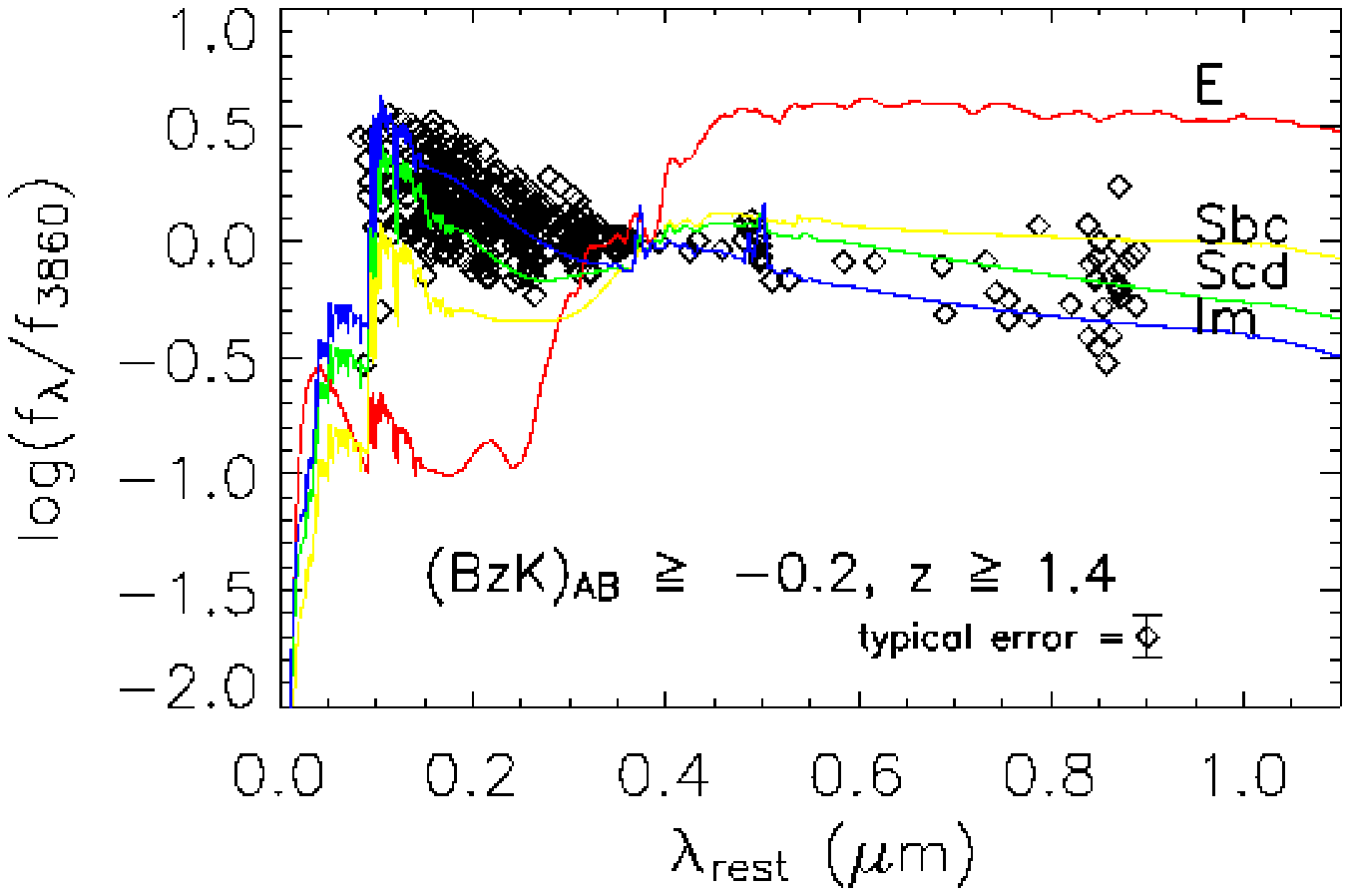}
\includegraphics[width=0.45\textwidth]{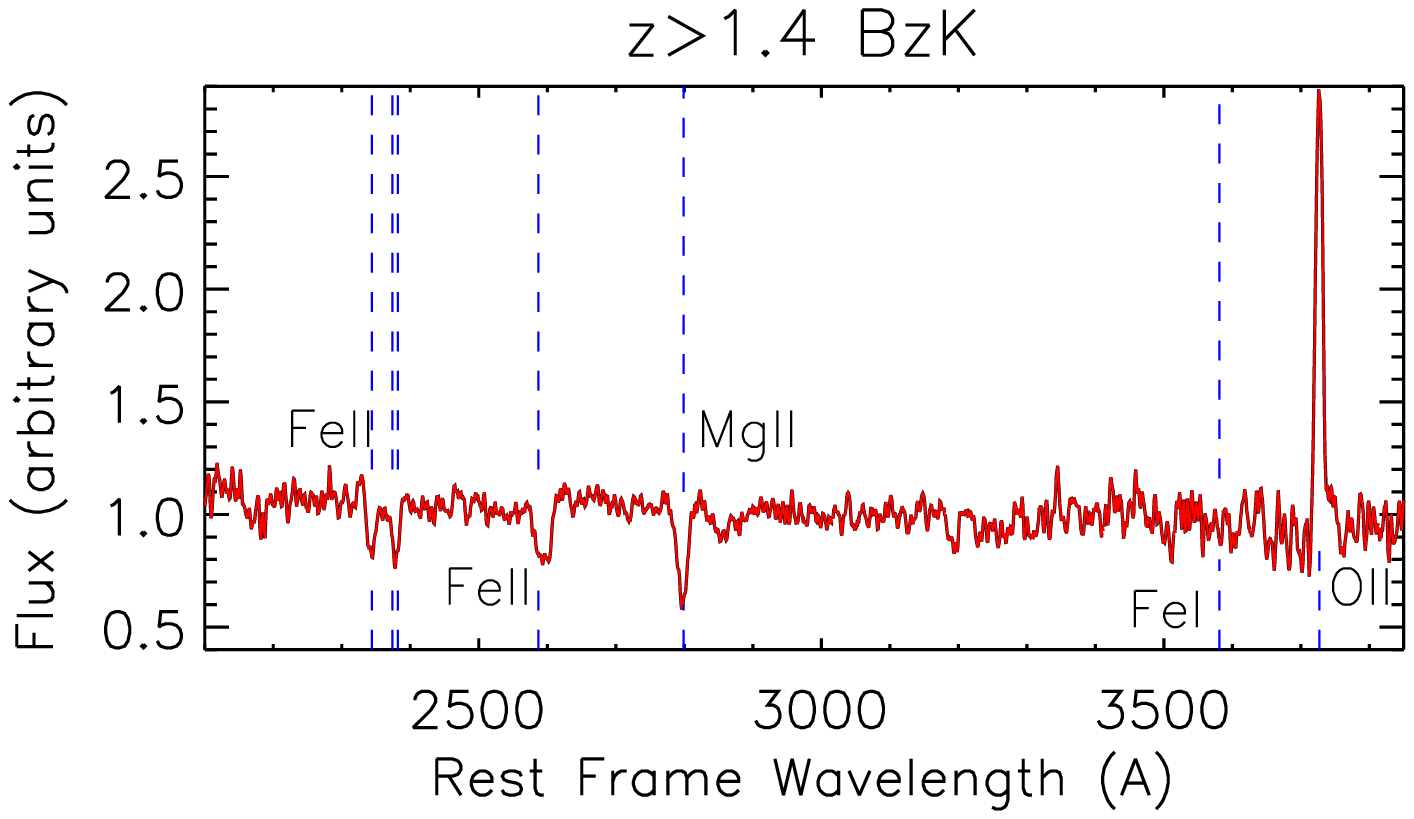}}
\caption{\emph{Left.} Rest-frame observed SEDs of z $>$ 1.4 galaxies
selected for having BzK $>$ $-$0.2. \emph{Right.} Composite spectrum
of the same galaxies.}
\label{fig:bzkcompo}
\end{figure}

The observed SEDs of objects with redshift $>$ 1.4 and BzK $>$ $-$0.2 closely 
approaches \cite{cww80} templates of late-type galaxies, in agreement with the 
expectations for this color selection. 
The composite spectrum obtained for these high redshift sources 
(Fig.~\ref{fig:bzkcompo}) 
actually shows the typical features of a late-type star forming galaxy.

\begin{acknowledgments}
This research has been developed within the framework of the VVDS
consortium.
This work has been partially supported by the
CNRS-INSU and its Programme National de Cosmologie (France),
and by Italian Ministry (MIUR) grants
COFIN2000 (MM02037133) and COFIN2003 (no.2003020150).
The VLT-VIMOS observations have been carried out on guaranteed
time (GTO) allocated by the European Southern Observatory (ESO)
to the VIRMOS consortium, under a contractual agreement between the
Centre National de la Recherche Scientifique of France, heading
a consortium of French and Italian institutes, and ESO,
to design, manufacture and test the VIMOS instrument.
\end{acknowledgments}

\bigskip

\noindent {\bf Complete author list}

\smallskip

\noindent S. Temporin$^1$, A. Iovino$^1$, H. J. McCracken$^{2,5}$, M. Bolzonella$^3$, M. Scodeggio$^4$, 
D. Bottini$^4$, B. Garilli$^4$, V. Le Brun$^6$, O. Le F\`evre$^6$, D. Maccagni$^4$,
J. P. Picat$^7$, R. Scaramella$^{8,10}$, L. Tresse$^6$, G. Vettolani$^8$, 
A. Zanichelli$^8$, C. Adami$^6$, S. Arnouts$^6$, S. Bardelli$^3$, A. Cappi$^3$,
S. Charlot$^{2,9}$, P. Ciliegi$^3$, T. Contini$^7$, O. Cucciati$^{1,14}$, S. Foucaud$^{21}$, P. Franzetti$^4$,
I. Gavignaud$^{11}$, L. Guzzo$^{1}$, O. Ilbert$^{20}$, B. Marano$^{12}$, C. Marinoni$^{18}$,
A. Mazure$^6$, B. Meneux$^{1,4}$, R. Merighi$^3$, S. Paltani$^{15,16}$, R. Pell\`o$^7$,
A. Pollo$^{6,17}$, L. Pozzetti$^3$, M. Radovich$^{13}$, G. Zamorani$^3$, E. Zucca$^3$,
M. Bondi$^8$, A. Bongiorno$^{12}$, J. Brinchmann$^{19}$, S. de la Torre$^6$,
F. Lamareille$^3$, Y. Mellier$^{2,5}$, P. Merluzzi$^{13}$, D. Vergani$^4$, C. J. Walcher$^6$
\medskip

\noindent $^1$INAF-Osservatorio Astronomico di Brera, Via Brera 28, 
20121 Milano, Italy \break email: giovanna.temporin@brera.inaf.it\\[\affilskip]
$^2$Institut d'Astrophysique de Paris, UMR 7095, 98 bis Bvd Arago, 75014 Paris, France\\[\affilskip]
$^3$INAF-Osservatorio Astronomico di Bologna, Via Ranzani 1, Bologna, Italy 
\\[\affilskip]
$^4$INAF-IASF, Via Bassini 11, Milano, Italy \\[\affilskip]
$^5$Observatoire de Paris, LERMA, 61 Avenue de l'Observatoire, 75014 Paris, 
France\\[\affilskip]
$^6$Laboratoire d'Astrophysique de Marseille, UMR 6110 CNRS-Universit\'e de
Provence,  BP8, 13376 Marseille Cedex 12, France\\[\affilskip]
$^7$Laboratoire d'Astrophysique de l'Observatoire Midi-Pyr\'en\'ees (UMR 
5572) - 14, avenue E. Belin, F31400 Toulouse, France\\[\affilskip]
$^8$IRA-INAF - Via Gobetti,101, I-40129, Bologna, Italy\\[\affilskip]
$^9$Max Planck Institut fur Astrophysik, 85741, Garching, Germany\\[\affilskip]
$^{10}$INAF-Osservatorio Astronomico di Roma - Via di Frascati 33,
I-00040, Monte Porzio Catone, Italy\\[\affilskip]
$^{11}$Astrophysical Institute Potsdam, An der Sternwarte 16, D-14482
Potsdam, Germany\\[\affilskip]
$^{12}$Universit\`a di Bologna, Dipartimento di Astronomia - Via Ranzani,1,
I-40127, Bologna, Italy\\[\affilskip]
$^{13}$INAF-Osservatorio Astronomico di Capodimonte - Via Moiariello 16, I-80131, Napoli,
Italy\\[\affilskip]
$^{14}$Universit\'a di Milano-Bicocca, Dipartimento di Fisica - 
Piazza delle Scienze, 3, I-20126 Milano, Italy\\[\affilskip]
$^{15}$Integral Science Data Centre, ch. d'\'Ecogia 16, CH-1290 Versoix\\[\affilskip]
$^{16}$Geneva Observatory, ch. des Maillettes 51, CH-1290 Sauverny, Switzerland\\[\affilskip]
$^{17}$Astronomical Observatory of the Jagiellonian University, ul Orla 171, 
30-244 Krak{\'o}w, Poland\\[\affilskip]
$^{18}$Centre de Physique Th\'eorique, UMR 6207 CNRS-Universit\'e de Provence, 
F-13288 Marseille France\\[\affilskip]
$^{19}$Centro de Astrofísica da Universidade do Porto, Rua das Estrelas,
4150-762 Porto, Portugal \\[\affilskip]
$^{20}$Institute for Astronomy, 2680 Woodlawn Dr., University of Hawaii,
Honolulu, Hawaii, 96822\\[\affilskip]
$^{21}$School of Physics \& Astronomy, University of Nottingham, University Park, Nottingham, NG72RD, UK\\[\affilskip]


\begin{thebibliography}{}

\bibitem[Bertin and Arnouts (1996)]{ba96}
     {Bertin, E. \&\ Arnouts, S.} 1996
     \textit{A\&AS} 117, 393

\bibitem[Coleman, Wu, and Weedman (1980)]{cww80}
     {Coleman, G. D., Wu, C.-C., and Weedman, D. W.} 1980
     \textit{ApJS} 43, 393

\bibitem[Daddi \etal\ (2004)]{Daddi04}
     {Daddi, E., Cimatti, A., Renzini, A. \etal} 2004,
     \textit{ApJ} 617, 746

\bibitem[Ilbert \etal\ (2006)]{oi06}
     {Ilbert, O., Arnouts, S., McCracken, H. J. \etal} 2006,
     \textit{A\&A} 457, 841
     
\bibitem[Iovino \etal\ (2005)]{ai05}
     {Iovino, A., McCracken, H. J., Garilli, B. \etal} 2005,
     \textit{A\&A} 442, 423       

\bibitem[Le F\'evre \etal\ (2005)]{olf05}
     {Le F\'evre, O., Vettolani, G., Garilli, B. \etal} 2005,
     \textit{A\&A} 439, 845
     
\bibitem[McCracken \etal\ (2003)]{hjmcc03}
     {McCracken, H. J., Radovich, M., Bertin, E. \etal} 2003,
     \textit{A\&A} 410, 17
     
\bibitem[Temporin \etal\ (2007a)]{st07a}
     {Temporin, S., Iovino, A., McCracken, H. J., and the VVDS Team} 2007a,
     \textit{A\&A}  in preparation
     
\bibitem[Temporin \etal\ (2007b)]{st07b}
     {Temporin, S., Iovino, A., McCracken, H. J., and the VVDS Team} 2007b,
     \textit{A\&A} in preparation
     
\end{thebibliography}
\end{document}